\documentclass[10pt]{article}

\usepackage{amsmath,amssymb}
\newcommand{\Act}{\mathcal{A}}
\newcommand{\Lagr}{\mathcal{L}}
\newcommand{\Wq}{\mathcal{W}}
\newcommand{\op}[1]{\hat{#1}}
\newcommand{\ket}[1]{| #1 \rangle}
\newcommand{\bra}[1]{\langle #1 |}

\newcommand{\el}{\mathsf l}

\newcommand{\mi}{\mathnormal i}
\newcommand{\me}{\mathnormal e}
\newcommand{\md}{\mathnormal d}

\newcommand{\arx}[1]{{\em Preprint} {\tt #1}}
\newcommand{\pra}{{\em Phys.\ Rev.\ A }}
\newcommand{\prl}{{\em Phys.\ Rev.\ Lett.\ }}

\newcommand{\Eq}[1]{Eq.~(\ref{#1})}
\begin{document}

\title{Towards Lagrangian approach to quantum computations (revised)}

\author{Alexander Yu.~Vlasov
}

\date{\today}
\maketitle

\begin{abstract}
 In this work is discussed possibility and actuality of Lagrangian approach
to quantum computations. Finite-dimensional Hilbert spaces used in this area
provide some challenge for such consideration. The model discussed here can be
considered as an analogue of Weyl quantization of field theory via path 
integral in L. D. Faddeev's approach. Weyl quantization is possible to use also
in finite-dimensional case, and some formulas may be simply rewritten with 
change of integrals to finite sums. On the other hand, there are specific 
difficulties relevant to finite case. 
\end{abstract}


\section{Introduction}

One initial reason to consider possibility of Lagrangian formalism in quantum
computations was an idea to use analogues of {\em a minimal
action principle} in some quantum optimization algorithms 
({\em cf\/} Ref.~\cite{Bennett}).  

In quantum physics the minimal action principle has standard
correspondence with classical one: any possible 
trajectory $\mathcal{P}$ of some particle has contribution 
$\exp(\frac{\mi}{\hbar}\int_\mathcal{P} \Lagr(t) \md t)$,
where $\Lagr$ is Lagrange function. Such an expression has an oscillatory 
behavior, and only a trajectory near extremum of action 
$\Act = \int_\mathcal{P} \Lagr(t) \md t$ 
matters, because other paths compensate each other due to interference
({\em cf\/} Ref.~\cite{QED}). 

Does it possible to use such principle in the quantum information science for 
finding an extremum of some function using a quantum mechanical system with 
appropriate Lagrangian? The idea encounters specific obstacles because Hilbert 
spaces used in quantum computations are finite-dimensional. On the other hand 
some analogue of Lagrangian theory may be really built and the present paper 
describes some basic properties of such models. In Section \ref{Cont} are 
briefly revisited ideas and formulas used in continuous case, and necessary for
a further revision with finite-dimensional Hilbert spaces discussed in Section 
\ref{Fin}. A relevance to the theory of quantum computations is discussed in 
Section \ref{Disc}.

\section{Continuous case revisited}
\label{Cont}
\subsection{Weyl quantization}

Let us recall briefly the idea of Weyl quantization \cite{WeylGQM} with most
attention to topics necessary for further applications to finite-dimensional 
Hilbert spaces.

In continuous case Weyl quantization uses \cite{WeylGQM,Fadd} a function $f(p,q)$ 
with two real arguments $p,q$ and with Fourier co-image $\tilde f(\alpha,\beta)$ 
described by expression
\begin{equation}
 f(p,q) = \iint{\exp(\mi\alpha p + \mi\beta q)
 \tilde f(\alpha,\beta)\, \md\alpha\, \md\beta}.
\label{fpq}
\end{equation}
Such a function is associated with an operator $\op f$ defined as
\begin{equation}
 \op f = \iint{\exp(\mi\alpha \op p + \mi\beta \op q)
 \tilde f(\alpha,\beta)\, \md\alpha\, \md\beta}.
\label{opf}
\end{equation}
Here $\op p$ and $\op q$ are {\em momentum} and {\em coordinate} operators.
In exponential form used in \Eq{opf} it is defined a {\em Weyl system},
{\em i.e.}, two families of operators:
\begin{equation}
 \op U(\alpha) = \exp(\mi\alpha \op p), \quad
 \op V(\beta) = \exp(\mi\beta \op q)
\end{equation}
satisfying relation \cite{WeylGQM} (in system of units with $\hbar=1$)
\begin{equation}
\op U(\alpha)\op V(\beta) = \exp(\mi\alpha\beta)\op V(\beta)\op U(\alpha),
\label{Wcom}
\end{equation}
equivalent to usual Heisenberg relation for coordinate and momentum
\begin{equation}
 [\op p,\op q] = \op p\op q -\op q\op p = -\mi.
\label{Hcom}
\end{equation}
Formally \Eq{Wcom} follows from \Eq{Hcom} due to a Campbell-Hausdorff formula
$$
\exp(\op a + \op b) = \exp(\op a)\exp(\op b)\exp(-\frac12[\op a,\op b])\exp R_2,
$$
where $R_2$ is a term with commutators of higher order, $R_2(\op p,\op q) =0$.
Using this formula it is possible also to rewrite \Eq{opf} as
\begin{equation}
 \op f = \iint{\tilde f(\alpha,\beta)\exp(\mi\alpha\beta/2)
 \op U(\alpha)\op V(\beta)\, \md\alpha\, \md\beta}.
\label{opfUV}
\end{equation}
 
Weyl quantization via \Eq{fpq} and \Eq{opf} produces a method of construction
of some operator $\op f$ for any function $f(p,q)$ with two variables
$
 \Wq: f \longrightarrow \op f.
$
It is enough to use the inverse Fourier transform in accordance with \Eq{fpq} to
create $\tilde f$ and use it for construction of the operator $\op f$.

There is also direct formula for $\op f$. 
Using an expression for kernel of \Eq{opf}
in coordinate representation \cite{Fadd}
\begin{equation}
\bra{q'}\me^{\mi\alpha \op p + \mi\beta \op q}\ket{q''} =
 \exp\bigl(\mi\frac{q'+q''}{2}\beta\bigr)\,\delta(q''-q'+\alpha),
\label{kernUV}
\end{equation}
it is possible to write elements of the operator $\op f$ \cite{Fadd}
\begin{equation}
\bra{q'}\op f\ket{q''} =
 \frac{1}{2\pi}\int{f\left(p,\frac{q'+q''}{2}\right)\me^{\mi p(q''-q')}\md p}.
\label{bfk}
\end{equation}
It is simple to invert Fourier transformation on $p$ used in \Eq{bfk} and 
using notation $q=(q''+q')/2$, $x = q''-q'$ to write
\begin{equation}
 f(p,q) = \int{\bra{q-x/2}\op f \ket{q+x/2} \exp(-\mi p x)\, \md x}.
\label{InvW}
\end{equation}
The formula \Eq{InvW} show, how to calculate $\Wq^{-1}$. 

It should be mentioned, 
that for a particular case, then $\op f$ is the statistical operator 
(``density matrix'') $\op \rho$, \Eq{InvW} coincides with an expression for 
{\em Wigner function}.
The property is known also as {\em Wigner-Weyl isomorphism} \cite{wwi}.
Despite of such coincidence in mathematical expressions, Weyl quantization has 
rather different area of applications. Wigner function was suggested for presentation 
of mixed states, but Weyl quantization often used for description of evolution of 
arbitrary operator in Heisenberg picture without actual necessity to work with 
states and density matrices. 

The Hamiltonian function here $H(p,q) = \Wq^{-1}\op H$ displays a proper 
correspondence principle with the classical physics, {\em i.e.}, the evolution
law for operators in the Heisenberg picture applied to coordinate and 
momentum
\begin{equation}
\frac{\md\op q}{\md t} = \mi [\op H,\op q], \quad
\frac{\md\op p}{\md t} = \mi [\op H,\op p],
\end{equation}
produces canonical Hamiltonian equations for classical coordinate,
momentum and function $H(p,q)$ \cite{WeylGQM}
\begin{equation}
\frac{\md q}{\md t} =  \frac{\partial H(p,q)}{\partial p}, \quad
\frac{\md p}{\md t} =  -\frac{\partial H(p,q)}{\partial q}.
\end{equation}
In continuous case such property ensures correspondence principle, because any
operator may be presented formally via series with $\op p$ and $\op q$. In the 
discrete case there are no good analogues of such operators and so other ideas 
should be used instead.

\subsection{Lagrangian approach and path integral}

The functional (path) integral in Weyl approach may be introduced by quite 
straightforward procedure \cite{Fadd}. In Schr\"odinger picture 
$\ket{\psi'} = \op S\ket{\psi}$ and it is possible to write elements of evolution
operator $\op S$ (``quantum gate'') for small period of time $\Delta t$ as
\begin{equation}
\bra{q''}\op S_{(\Delta t)}\ket{q'} =
\bra{q''}\me^{-\mi\op H\Delta t}\ket{q'} \cong
\bra{q''}1 -\mi\op H\Delta t\ket{q'}.
\label{Sqq}
\end{equation}
Applying \Eq{bfk} to \Eq{Sqq}
\begin{eqnarray}
\bra{q''}\op S\ket{q'} &\cong&
\frac{1}{2\pi}\int \me^{\mi p(q''-q')}\bigl(1-\mi H(p,\tfrac{q''+q'}{2})\Delta t\bigr) \md p
\nonumber\\
&\cong&
\frac{1}{2\pi}\int \me^{\mi p(q''-q')-\mi H(p,\frac{q''+q'}{2})\Delta t} \md p
\label{Sdt}
\end{eqnarray}
It is possible to divide finite interval of time on $N$ small periods 
and use expression $\op S = \prod \exp(\mi \op H(t)\Delta t)$.
For $N \to \infty$ and $\Delta t \to 0$ such products of \Eq{Sdt} corresponds 
to expression for functional (continual, path) integral along all paths in 
coordinate and momentum space
\begin{equation}
 \bra{q''}\op S\ket{q'} =
 \int_\mathcal{P}\exp\Bigl(\mi {\textstyle\int\bigr(p \dot q - H(p,q)\bigl)dt}\Bigr)
 \mathcal{D}p\mathcal{D}q.
\label{fuctint}
\end{equation}
The function $\Act = \int\bigr(p \dot q - H(p,q)\bigl)\md t$ here is limit 
$\Delta t\to 0$ of finite sums in \Eq{Sdt} and coincides with {\em classical 
action}. More details about finite sums used in the passage to the continuous 
limit may be found also in description of \Eq{PathSum} below.

\section{Finite-dimensional case}
\label{Fin}
\subsection{Weyl pair}

An advantage of Weyl relations \Eq{Wcom} for purposes of present paper
is possibility to write an analogue of such operators in finite, 
$n$-dimensional case \cite{WeylGQM}. It is enough to use Weyl pair of 
$n \times n$ matrices $\op U$, $\op V$:
\begin{equation}
 \op U_{jk} = \delta_{j,k+1\bmod n}, \quad
 \op V_{jk} = \exp\bigl(\frac{2\pi\mi}{n}j\bigr)\delta_{jk}
\end{equation}
with property
\begin{equation}
\op U\op V = \exp(2\pi\mi /n)\op V\op U.
\label{UVVU}
\end{equation}

Let us consider finite analogues of other expressions.
Let $\tilde M(a,b)$ is function of two integer arguments $a,b = 0,\ldots,n-1$.
It is possible to use discrete Fourier transform for both arguments and write
analogue of \Eq{fpq}
\begin{equation}
 M(p,q) = \frac{1}{n}\sum_{a,b=0}^{n-1}%
 \exp\left(\frac{2\pi\mi}{n}(ap + bq)\right)\tilde M(a,b).
\label{Mpq}
\end{equation}
The function $M(p,q)$ is defined for any real $p,q$, but integer values are 
enough to find $\tilde M(a,b)$ via inverse Fourier transform, and so value of 
$M(p,q)$ for any real values.  

Analogue of operator $\op f$ defined by \Eq{opfUV} is $n \times n$ matrix 
$\op M$ could be written as 
\begin{equation}
 \op M = \frac{1}{n}\sum_{a,b=0}^{n-1}%
 \exp\left(\frac{\pi\mi}{n}\,ab\right)\op U^a \op V^b\tilde M(a,b),
\label{opM}
\end{equation}
but it produces some problems due to using of modular arithmetics,
{\em e.g.} unlike with continuous case the operator \Eq{opM} $\op M$ 
is not Hermitian for real function $M(p,q)$ \Eq{Mpq}. 
Anyway, for simpler illustration of suggested
approach, it is enough to consider asymmetric construction, {\em i.e.},
\begin{equation}
 \op M^\el = \frac{1}{n}\sum_{a,b=0}^{n-1}\op U^a \op V^b\tilde M(a,b),
\label{opMl}
\end{equation}

For such operator true discrete {\em asymmetric} analogue of expression \Eq{bfk}
\begin{equation}
\op M^\el_{kj} =
\bra{j}\op M^\el\ket{k} = \frac{1}{n}\sum_{p=0}^{n-1}%
\me^{\frac{2\pi\mi}{n}p(j-k)}M(p,k).
\label{bMka}
\end{equation}
It can be checked directly using \Eq{Mpq} and \Eq{opMl}:
\begin{eqnarray*}
\lefteqn{
\frac{1}{n}\sum_{p=0}^{n-1}\me^{\frac{2\pi\mi}{n}p(k-j)}M(p,k) = 
 \frac{1}{n}\sum_{p,a,b=0}^{n-1}\me^{\frac{2\pi\mi}{n}(p(k-j)+ap+bk)}\tilde M(a,b)}\\
&&=\frac{1}{n}\sum_{a,b=0}^{n-1}\!\me^{\frac{2\pi\mi}{n}bk}\tilde M(a,b)%
\underbrace{\sum_{p=0}^{n-1}\me^{\frac{2\pi\mi}{n}p(k-j+a)}}_{\textstyle n\delta_{j-k,a}}
=\sum_{a,b=0}^{n-1}
\underbrace{\delta_{j-k,a}\me^{\frac{2\pi\mi}{n}bk}}_{\textstyle(\op U^a \op V^b)_{kj}}
\tilde M(a,b)
=\op M^\el_{kj}.
\end{eqnarray*}

Inverse transformation for \Eq{bMka} also may be simply found
\begin{equation}
M(p,q) = \sum_{j=0}^{n-1}\me^{-\frac{2\pi\mi}{n}pj}\op M^\el_{q,q+j},
\quad p,q \in \mathbb Z.
\label{iMka}
\end{equation}
It should be mentioned also, that in equations above notation for modular arithmetic
is often omitted for simplicity, {\em e.g.}, $q+j$ used instead of $q+j \mod n$.

\subsection{Lagrangian approach to discrete models}

Using \Eq{bMka} it is possible to write analogue of Lagrangian function \Eq{Sdt} 
\begin{eqnarray}
\bra{j}\op S^\el\ket{k} &\cong&
\frac{1}{n}\sum_{p=0}^{n-1}%
 \me^{\frac{2\pi\mi}{n}p(j-k)}\bigl(1-\mi H(p,j)\Delta t\bigr)
\nonumber\\
&\cong&
\frac{1}{n}\sum_{p=0}^{n-1}\me^{\frac{2\pi\mi}{n}p(j-k)-\mi H(p,j)\Delta t}.
\label{SumDt}
\end{eqnarray}
Here $\bra{j}\op S\ket{k} = \op S_{kj}$ are simply indexes of matrix. For 
interval of time divided on $N$ segments we have
\begin{equation}
\bra{q_0}\op S\ket{q_N} = \!\! \sum^{n-1}_{q_1,\ldots,q_{N-1}=0}\!\! 
 \bra{q_0}\op S_1\ket{q_1}
 \cdots \bra{q_{N-1}}\op S_N\ket{q_N}.
\label{MatrPath}
\end{equation}
Using \Eq{SumDt} it is possible to write
\begin{equation}
\bra{q_0}\op S^\el\ket{q_N} \cong \frac{1}{n^N}
 \sum^{n-1}_{\substack{q_1,\ldots,q_{N-1}=0\\p_0,\ldots,p_{N-1}=0}} 
\exp\bigl(\mi\mathfrak{A}^\el_{q,p}\bigr)
\label{PathSum}
\end{equation}
where discrete analogue of action
\begin{eqnarray}
\mathfrak{A}^\el_{q,p} &=&
\sum_{k=0}^{N-1}\bigl( \frac{2\pi}{n}p_k (q_{k+1}-q_k)
 -H(p_k,q_k)\Delta t\bigr)
\nonumber \\
&=& \sum_{k=0}^{N-1}\bigl( \frac{2\pi}{n}p_k \frac{\Delta q_k}{\Delta t}
 -H(p_k,q_k)\bigr)\Delta t
\label{DiscrAct}
\end{eqnarray}
is calculated along all $n^{N-1}$ possible paths $q_k$ between two fixed points 
and with $n^N$ different momentum $p_k$ for each separate segment of such
broken line. 
Approximate expresion \Eq{PathSum} converges\footnote{Strictly speaking, such 
limit for path integrals is usually not well-defined.} to value 
$\op S^\el_{q_0q_N}$ in limit \mbox{$N \to \infty$}, \mbox{$\Delta t \to 0$}.
For continuous limit $n \to \infty$ the \Eq{DiscrAct} for asymmetric ordering may 
look even more traditional \cite{Wein}.

\subsection{Precise expression with ``effective Lagrangian''}

For further analysis of discrete models it is possible to introduce some
``effective Lagrangian'' 
to have precise value of $\op S^\el_{jk}$
for any division of time interval, including $N=1$. Really, errors in 
expressions above was related with two consequent approximations like
$\exp(\mi\epsilon) \cong 1 + \mi\epsilon$.  
Let us instead of application of (discrete) Weyl quantisation to Hamiltonian 
apply it directly to operator $\op S^\el$ in \Eq{SumDt}. 
\begin{eqnarray}
\bra{j}\op S^\el\ket{k} &=&
\frac{1}{n}\sum_{p=0}^{n-1}\me^{\frac{2\pi\mi}{n}p(j-k)}S(p,j)
\nonumber\\
&=& \frac{1}{n}\sum_{p=0}^{n-1}\me^{\frac{2\pi\mi}{n}p(j-k) + \ln S(p,j)}
\label{efSumDt}
\end{eqnarray}
with straightforward generalization to sum like \Eq{PathSum}, precise for any 
$N \geqslant 1$. It is only necessary to change $H$ to 
$H'_{\Delta t} \equiv \mi\ln S_{(\Delta t)} / \Delta t$, it 
is an ``effective Hamiltonian,'' $H'_{\Delta t} \to H$, $\Delta t \to 0$.
Here all paths with terms $S(p,q)=0$ should be simply omitted.

\subsection{Discrete Fourier transform}

For some cases the precise expression \Eq{efSumDt} may be quite simple.
Let us consider for example discrete Fourier transform
\begin{equation}
 \op F_{jk} = \frac{1}{\sqrt{n}}\exp\bigl(\frac{2\pi\mi}{n}jk\bigr).
\end{equation}

Using inverse transformation \Eq{iMka} it is possible to find
\begin{equation}
 F(p,q) = \frac{1}{\sqrt{n}}\sum_{j=0}^{n-1}\me^{\frac{2\pi\mi}{n}[q(q+j)-pj]}
= \sqrt{n}\,\me^{\frac{2\pi\mi}{n}q^2}\delta_{pq}
= \sqrt{n}\,\me^{\frac{\pi\mi}{n}(p^2+q^2)}\delta_{pq}.
\label{effF}
\end{equation}

Here ``effective Hamiltonian'' $q^2$ was rewritten in symmetric form 
\mbox{$(q^2 + p^2)/2$} due to term $\delta_{pq}$. It is convenient also, because
for quantum computation with continuous variables the Hamiltonian of
harmonic oscillator really may be used for realization of Fourier 
transform \cite{contvar}.

\section{Discussion}
\label{Disc}

In general, quantum computation is theory about efficient solving of computationally
hard problems using quantum systems and processes. Difficulty of computational
problems relevant with Lagrangian approach and calculation of sums over paths 
has two reasons. First one is usual exponential growth of Hilbert space dimension 
with respect to number of quantum systems, {\em i.e.}, calculations with $n$ qubits 
requires  operations with $2^n \times 2^n$ matrices (multiplications, exponents, 
{\em etc.}). The second one is well-known difficulty with definition and calculation of
functional (path) integrals, it may be described as unspecified growth of number of 
terms in sums like \Eq{PathSum} due to using expressions with limits $\Delta t \to 0$, 
$N \to \infty$. 

\smallskip

It was already mentioned some mathematical resemblance with theory of Wigner
functions in continuous case. 
For discrete models some methods from theory of Wigner function like doubling 
of lattice, using reflection operator and Galois fields \cite{MPS,Wootters} do not 
have simple extensions for theory considered here, but maybe it provides
promising challenge. Anyway, it is quite likely, that due to some analogy 
between theory of Weyl quantization and Wigner function, it is possible to 
combine such techniques for particular tasks related with statistical description 
of ensembles in theory of quantum computations and communications.

\smallskip

In considered theory Fourier transform is related with simplest quadratic
Hamiltonian similar with harmonic oscillator.
In usual, continuous theory only quadratic Hamiltonians produce possibility of 
more or less rigor calculation of path integral. In discrete theory used 
here all expressions may be calculated for any functional dependence, but 
simple representation of Fourier transformation may be convenient, 
because it is a fundamental tool for many quantum algorithms
\cite{JozF} including Shor's factoring one \cite{Shor}. 

\smallskip
 
It is also possible to compare such approach with yet another propositions
for quantum algorithms, despite they may look different from traditional
one. For example it was suggested \cite{Cer} to use many-slits interference
with different paths for resolution of NP-complete problems on quantum computer.

\smallskip

It should be mentioned also idea of quantum optimization algorithms discussed
in introduction. It works, if due to oscillatory behaviour and appropriate
Lagrangian, all paths except optimal are vanishing. It is usual for 
classical limit. 

On the other hand, the method may be useful for calculation of {\em sum itself}, 
even if it is not related directly with any optimization task. For example, 
application of sum over paths for computing of polynomial equation over finite 
fields was discussed recently in \cite{DHHMNO}. It was also noted there,
that many basic papers about quantum computing complexity use some
variants of sum over path approach (see \cite{DHHMNO} and references therein).

Some difference between the just mentioned models and the method considered
in this paper similar with distinction between earlier version with sum other 
{\em spatial} paths and more recent version with paths in the phase $(p,q)$ space 
\cite{Fadd,Wein}.

The models discussed in \cite{DHHMNO} exploit straightforward representation
of matrix multiplication via path summation \Eq{MatrPath} with only one kind of 
indexes $(q_k)$, but in present paper is also used more general  expressions with 
sum other paths in whole discrete phase space $(p_k,q_k)$ \Eq{PathSum}.
Only for some particular cases, like quadratic Hamiltonians or ``effective Hamiltonian''
with $\delta$ symbol \Eq{effF}, the summation on $p_k$ may be cancelled.  

The detailed representation in phase space is essential, because one purpose of 
this paper was to find relations between physical models and abstract quantum 
computations, {\em e.g.}, it becomes clearer from equations considered above, 
how Lagrangian term $p\Delta q - H\Delta t$ appears in expressions for 
sum over paths, how classical limit may be obtained for quantum 
system due to growth of Hilbert space dimension, {\em etc.} 

\smallskip

There are also other tasks relevant to given approach. For example, Lagrangian 
formalism is standard method for description of interacting quantum fields
and it may be useful for more adequate models of quantum gates and computation.

\section*{Acknowledgements}
Author is grateful to Seth Lloyd for encouragement and drawing attention 
to discrete lattice models.

\end{document}